\documentclass[12pt,twoside]{article}
\usepackage{a4}

\begin{document}

\newcommand{\gf}{G_{\mbox{{\scriptsize F}}}}
\newcommand{\order}{{\cal O}}
\newcommand{\delp}{\Delta P}
\newcommand{\leqsim}{\,\mbox{{\scriptsize $\stackrel{<}{\sim}$}}\,}
\newcommand{\geqsim}{\,\mbox{{\scriptsize $\stackrel{>}{\sim}$}}\,}
\newcommand{\bskk}{B_s\to K^+ K^-}
\newcommand{\bsokk}{B_s^0\to K^+ K^-}
\newcommand{\bsobkk}{\bar B_s^0\to K^+ K^-}
\newcommand{\bppipko}{B^+\to \pi^+ K^0}
\newcommand{\bmpimkob}{B^-\to \pi^- \bar K^0}
\newcommand{\bmpiokm}{B^-\to \pi^0 K^-}
\newcommand{\bppiokp}{B^+\to \pi^0 K^+}
\newcommand{\bdpimkp}{B^0_d\to \pi^- K^+}
\newcommand{\bdbpipkm}{\bar B^0_d\to \pi^+ K^-}
\newcommand{\pew}{P_{\mbox{{\scriptsize EW}}}}
\newcommand{\pewb}{\bar P_{\mbox{{\scriptsize EW}}}}
\newcommand{\pewp}{P_{\mbox{{\scriptsize EW}}}'}
\newcommand{\pewc}{P_{\mbox{{\scriptsize EW}}}^{\mbox{{\scriptsize C}}}}
\newcommand{\pcew}{P_{\mbox{{\scriptsize EW}}}^{\mbox{{\scriptsize (C)}}}}
\newcommand{\pcewb}{\bar P_{\mbox{{\scriptsize
EW}}}^{\mbox{{\scriptsize (C)}}}}
\newcommand{\pewpc}{P_{\mbox{{\scriptsize EW}}}'^{\mbox{{\scriptsize C}}}}
\newcommand{\bbtodb}{\bar b\to\bar d}
\newcommand{\bbtosb}{\bar b\to\bar s}

\newcommand{\bkk}{B_d\to K^0\bar K^0}
\newcommand{\bdkk}{B_d^0\to K^0\bar K^0}
\newcommand{\bdbkk}{\bar B_d^0\to K^0\bar K^0}
\newcommand{\bpipi}{B_d\to\pi^+\pi^-}
\newcommand{\bdpipi}{B_d^0\to\pi^+\pi^-}
\newcommand{\bdbpipi}{\bar B_d^0\to\pi^+\pi^-}
\newcommand{\bksks}{B_d\to K_{\mbox{{\scriptsize S}}}
K_{\mbox{{\scriptsize S}}}}
\newcommand{\bpiphi}{B_s\to\pi^0\Phi}
\newcommand{\bopiphi}{B_s^0\to\pi^0\Phi}
\newcommand{\bbopiphi}{\bar B_s^0\to\pi^0\Phi}
\newcommand{\brhok}{B_s\to\rho^0K_{\mbox{{\scriptsize S}}}}
\newcommand{\bqf}{B_q\to f}
\newcommand{\pcps}{\phi_{\mbox{{\scriptsize CP}}}(B_s)}
\newcommand{\pcpq}{\phi_{\mbox{{\scriptsize CP}}}(B_q)}
\newcommand{\pw}{\phi_{\mbox{{\scriptsize W}}}}
\newcommand{\acp}{a_{\mbox{{\scriptsize CP}}}}
\newcommand{\acpdir}{{\cal A}_{\mbox{{\scriptsize CP}}}^
{\mbox{{\scriptsize dir}}}}
\newcommand{\acpmi}{{\cal A}_{\mbox{{\scriptsize
CP}}}^{\mbox{{\scriptsize mix-ind}}}}
\newcommand{\acc}{A_{\mbox{{\scriptsize CC}}}}
\newcommand{\aew}{A_{\mbox{{\scriptsize EWP}}}}
\newcommand{\heff}{{\cal H}_{\mbox{{\scriptsize eff}}}(\Delta B=-1)}
\newcommand{\xif}{\xi_f^{(q)}}
\newcommand{\xiskk}{\xi_{K^+K^-}^{(s)}}
\newcommand{\xikk}{\xi_{K^0\bar K^0}^{(d)}}
\newcommand{\VmA}{\mbox{{\scriptsize V--A}}}
\newcommand{\VpA}{\mbox{{\scriptsize V+A}}}
\newcommand{\VpmA}{\mbox{{\scriptsize V$\pm$A}}}
\newcommand{\beq}{\begin{equation}}
\newcommand{\eeq}{\end{equation}}
\newcommand{\bea}{\begin{eqnarray}}
\newcommand{\eea}{\end{eqnarray}}
\newcommand{\non}{\nonumber}
\newcommand{\lab}{\label}
\newcommand{\la}{\langle}
\newcommand{\ra}{\rangle}
\newcommand{\np}{Nucl.\ Phys.}
\newcommand{\pl}{Phys.\ Lett.}
\newcommand{\prl}{Phys.\ Rev.\ Lett.}
\newcommand{\pr}{Phys.\ Rev.}
\newcommand{\zp}{Z.\ Phys.}

\setcounter{page}{-1}
\vspace*{-1cm}
\thispagestyle{empty}
\begin{flushright}
MPI-PhT/95-72\\
TUM-T31-96/95\\
TTP95-30\\
hep-ph/9507460\\
July 1995
\end{flushright}

\begin{center}
\vspace{0.2cm}
{\Large{\bf A Determination of the CKM-angle $\alpha$ using}}\\
\vspace{0.2cm}
{\Large{\bf Mixing-induced CP Violation in the Decays}}\\
\vspace{0.2cm}
{\Large{\bf  $B_d\to\pi^+\pi^-$ and $B_d\to K^0\bar K^0$
\footnote[1]{Supported in part by the German {\it Bundesministerium
f\"ur Bildung und Forschung} under contract 06--TM--743 and
by the CEC science project SC1--CT91--0729.}}}\\
\vspace{0.7cm}
{\large{\sc Andrzej J. Buras}}\\

\vspace{0.1cm}
{\sl Technische Universit\"at M\"unchen, Physik Department\\
D--85748 Garching, Germany}\\
\vspace{0.3cm}
{\sl Max-Planck-Institut f\"ur Physik\\
-- Werner-Heisenberg-Institut --\\
F\"ohringer Ring 6, D--80805 M\"unchen, Germany}\\

\vspace{0.7cm}

{\large{\sc Robert Fleischer}}\\
\vspace{0.1cm}
{\sl Institut f\"ur Theoretische Teilchenphysik\\
Universit\"at Karlsruhe\\
D--76128 Karlsruhe, Germany}\\
\vspace{0.8cm}
{\large{\bf Abstract}}
\end{center}
\vspace{0.2cm}
We present a method of determining the CKM-angle $\alpha$ by performing
simultaneous measurements of the mixing-induced CP asymmetries of the
decays $B_d\to\pi^+\pi^-$ and $B_d\to K^0\bar K^0$. The accuracy of our
approach is limited by $SU(3)$-breaking effects originating from $\bar b\to
\bar ds\bar s$ QCD-penguin diagrams. Using plausible power-counting
arguments we show that these uncertainties are expected to be of the same
order as those arising through electroweak penguins in the standard
Gronau-London-method in which $\alpha$ is extracted by means of
isospin relations among $B\to\pi\pi$ decay amplitudes.
Therefore our approach, which does not involve the experimentally difficult
mode $B_d\to\pi^0\pi^0$ and is essentially unaffected by electroweak
penguins, may be an interesting alternative to determine $\alpha$.

\newpage
\thispagestyle{empty}
\mbox{}
\newpage

CP-violating asymmetries arising in nonleptonic $B$-decays
(see e.g.\ refs.~\cite{bs}-\cite{qui}) will play a central role in the
determination of the angles $\alpha$, $\beta$ and $\gamma$ in the
unitarity triangle \cite{ck,js} at future experimental $B$-physics projects.
Unfortunately, these
asymmetries are in general not related to the CKM-angles in a clean way,
but suffer from uncertainties originating from so-called penguins.
These contributions preclude in particlular a clean determination of the
CKM-angle $\alpha$ by measuring the mixing-induced CP-violating
asymmetry $\acpmi(B_d\to\pi^+\pi^-)$. In a pioneering paper \cite{gl},
Gronau and London have presented a method to eliminate the uncertainty
in this determination of $\alpha$ that is related to {\it QCD-penguins}.
It uses isospin relations among $B_d\to\pi^+\pi^-$, $B_d\to\pi^0\pi^0$ and
$B^\pm\to\pi^\pm\pi^0$ decay amplitudes and requires besides a
time-dependent study of $\bpipi$ yielding $\acpmi(\bpipi)$ a
measurement of the corresponding branching ratios.

However, there are not only QCD- but also {\it electroweak} penguin
operators. \mbox{Although} one would expect na\"\i vely that electroweak
penguins should only play a minor role in nonleptonic $B$-decays, there
are certain transitions that are affected significantly by these
operators which become important in the presence of a heavy top-quark.
This interesting feature has first been pointed out in
refs.~\cite{rfewp1}-\cite{rfewp3} and has been confirmed later by the
authors of refs.~\cite{dhewp1}-\cite{dy}. As has been stressed first by
Deshpande and He \cite{dhewp2}, the influence of electroweak penguins on
the extraction of $\alpha$ by using the standard
Gronau-London-method \cite{gl} could also be sizable. A more
elaborate analysis \cite{ghlrewp} shows, however, that this impact is
expected to be rather small, at most a few per cent.

In a recent paper \cite{bfewp} we have presented strategies for the
experimental determination of electroweak penguin contributions
to nonleptonic $B$-decays. These strategies allow in particular
to control the electroweak penguin uncertainty \mbox{affecting} the
extraction of the CKM-angle $\alpha$ in the Gronau-London-method \cite{gl}.
Although this method of determining $\alpha$ is very clean from the
theoretical point of view, it requires the measurement of the decay
$B_d\to\pi^0\pi^0$ which is rather difficult. The very recent analysis
by Kramer and Palmer \cite{kp} indicates a branching ratio
BR$(B_d\to\pi^0\pi^0)\leqsim\order(10^{-6})$. Therefore, it is important
to search for other methods that allow a clean determination of $\alpha$.
Such methods are also needed for overconstraining the shape of the
unitarity triangle.

Motivated by this experimental situation, Dunietz \cite{dunrev} has
suggested an alternative way of extracting $\alpha$ that is based on
the $SU(3)$ flavour symmetry of strong interactions \cite{zep}-\cite{hmt}
and uses time-dependent measurements of the modes $\bpipi$ and $\bskk$.
However in view of the large $B^0_s$--$\bar B^0_s$--mixing,  the
time-dependent analysis of the transition $\bskk$ with the expected
branching ratio at the $\order(10^{-5})$ level may be difficult
as well.

In this letter we would like to propose a different method of
extracting $\alpha$. In order to eliminate the penguin contributions, we
use time-dependent measurements of the modes $\bpipi$ and $\bkk$ yielding
the corresponding mixing-induced CP-violating asymmetries and
employ the $SU(3)$ flavour symmetry of strong interactions
\cite{zep}-\cite{hmt} to derive relations among the corresponding decay
amplitudes. The transition $\bkk$ is -- in contrast to $\bskk$ -- a pure
penguin-induced mode with a branching ratio $\order(10^{-6})$
\cite{kp,rfkokob}. Yet because
of smaller $B^0_d$--$\bar B^0_d$--mixing, time-dependent studies of this
channel may probably be easier for experimentalists than those of
the decay $\bskk$. As we will see in a moment, our approach is essentially
unaffected by electroweak penguins.

In the previous literature it has been claimed by several authors
that the Standard Model predicts {\it vanishing} CP-violating
asymmetries for decays such as $\bksks$ or $\bkk$ (the CP asymmetries
of both channels are equal) because of the cancellation of weak decay- and
mixing-phases (see e.g.\ refs.~\cite{lp,nir,qui}). This result is
however only correct, if the $\bar b\to\bar d$ QCD-penguin amplitudes are
dominated by internal top-quark exchanges. As has been pointed out in
refs.~\cite{rfkokob,bf}, QCD-penguins with internal up- and charm-quarks
may generally also play a significant role and in the case of $\bkk$ could
lead to rather large CP asymmetries of $\order(10-50)\%$ \cite{rfkokob}.
Unfortunately, these asymmetries suffer from large hadronic uncertainties
and are therefore not related to CKM-angles in a clean way. Nevertheless,
$\acpmi(\bkk)$ may be combined with additional inputs to determine
$\alpha$ in a clean way as we will demonstrate in this letter.

In our discussion it is convenient to use the description of
$B\to PP$ decays given by
Gronau, Hern\'andez, London and Rosner in refs.~\cite{ghlrewp} and
\cite{grl}-\cite{ghlrsu3}. Using the same notation as these authors,
the $\bdpipi$ and $\bdkk$ decay amplitudes take the form
\beq\lab{e1}
\begin{array}{rcl}
A(\bdpipi)&=&-\left[(T+E)+(P+PA)+c_u\pewc\right]\\
A(\bdkk)&=&\left[(P+PA+P_3)+c_s\pewc\right],
\end{array}
\eeq
where $T$ and $E$ describe $\bar b\to\bar uu\bar d$ colour-allowed
tree-level and exchange
amplitudes, respectively, $P$ denotes $\bbtodb$ QCD-penguins,
$PA$ is related to QCD-penguin annihilation diagrams and $\pewc$
to colour-suppressed $\bbtodb$ electroweak penguins. The term
$P_3$ describes $SU(3)$-breaking effects that are introduced through the
creation of a $s\bar s$ pair in the $\bbtodb$ QCD-penguin diagrams
\cite{ghlrsu3}. If we follow the plausible arguments of Gronau
et al.\ outlined in \cite{ghlrewp,ghlrsu3}, we expect the
following hierarchy of the different topologies present in (\ref{e1}):
\beq\lab{e2}
\begin{array}{rcl}
1&:&|T|\\
\order(\bar\lambda)&:&|P|\\
\order(\bar\lambda^2)&:&|E|, \quad \left|P_3\right|\\
\order(\bar\lambda^3)&:&|PA|, \quad \left|\pewc\right|.
\end{array}
\eeq
Note that the parameter $\bar\lambda=\order(0.2)$ appearing in these
relations is not related to the usual Wolfenstein parameter $\lambda$.
It has been introduced by Gronau et al.\ just to keep track of the
expected orders of magnitudes. We have named this quantity $\bar\lambda$
in order not to confuse it with Wolfenstein's $\lambda$.

Consequently, if we neglect the terms of $\order(\bar\lambda^3)$,
we obtain
\beq\lab{e3}
\begin{array}{rcl}
A(\bdpipi)&=&-\left[(T+E)+P\right]\\
A(\bdkk)&=&P+P_3.
\end{array}
\eeq
Within this approximation, terms of $\order(\bar\lambda^4)$, i.e.\
$SU(3)$-breaking corrections to the $PA$ and $\pewc$ amplitudes, which
have not been written explicitly in (\ref{e1}), have also to be neglected.

Rotating the $\bdbpipi$ and $\bdbkk$ amplitudes by the phase factor
$e^{-2i\beta}$, we find
\beq\lab{e4}
\begin{array}{rcl}
e^{-2i\beta}A(\bdbpipi)&=&-\left[e^{2i\alpha}(T+E)+
e^{-2i\beta}\bar P\right]\\
e^{-2i\beta}A(\bdbkk)&=&e^{-2i\beta}\left(\bar P+\bar P_3\right),
\end{array}
\eeq
where we have used the relation
\beq\lab{e5}
e^{-2i\beta}(\bar T+\bar E)=e^{-2i(\beta+\gamma)}(T+E)=e^{2i\alpha}(T+E).
\eeq
Using (\ref{e3}) and (\ref{e4}) it is an easy exercise to eliminate
$P$ and $\bar P$ and to derive the following relations:
\beq\lab{e6a}
A(\bdkk)+(T+E)-P_3+A(\bdpipi)=0
\eeq
\beq\lab{e6b}
e^{-2i\beta}A(\bdbkk)+e^{2i\alpha}(T+E)-
e^{-2i\beta}\bar P_3+e^{-2i\beta}A(\bdbpipi)=0,
\eeq
which have been represented graphically in the complex plane in Fig.~1.
If the $\bar b\to\bar d$ QCD-penguins were dominated by internal
top-quark exchanges, we would have $e^{-2i\beta}\bar P_3=P_3$. However,
as has been shown in refs.~\cite{rfkokob,bf}, QCD-penguins with internal
up- and charm-quarks are expected to lead to sizable corrections to this
relation.

The angles $\psi$ and $\phi$ appearing in Fig.~1 can be determined
{\it directly} by measuring the mixing-induced CP asymmetries of the
decays $\bkk$ and $\bpipi$, respectively, which are given by \cite{bfewp}
\bea
\acpmi(\bkk)&=&-\frac{2|A(\bdbkk)||A(\bdkk)|}{|A(\bdbkk)|^2+
|A(\bdkk)|^2}\sin\psi\lab{e7}\\
\acpmi(\bpipi)&=&-\frac{2|A(\bar B^0_d\to\pi^+\pi^-)|
|A(B^0_d\to\pi^+\pi^-)|}{|A(\bar B^0_d\to\pi^+\pi^-)|^2+
|A(B^0_d\to\pi^+\pi^-)|^2}\sin\phi\lab{e8}
\eea
and enter the formulae for the corresponding time-dependent CP asymmetries
in the following way:
\bea
\lefteqn{\acp(t)\equiv\frac{\Gamma(B_d^0(t)\to f)-\Gamma(\bar
B_d^0(t)\to f)}{\Gamma(B_d^0(t)\to f)+\Gamma(\bar B_d^0(t)\to f)}=}\lab{e9}\\
&&\acpdir(B_d\to f)\cos(\Delta M_d t)+
\acpmi(B_d\to f)\sin(\Delta M_d t).\nonumber
\eea
Here, $\acpdir(B_d\to f)$ describes direct CP violation and is given by
\beq\lab{e10}
\acpdir(B_d\to f)=\frac{|A(B^0_d\to f)|^2-|A(\bar B^0_d\to f)|^2}
{|A(B^0_d\to f)|^2+|A(\bar B^0_d\to f)|^2},
\eeq
whereas $\Delta M_d$ denotes the mass splitting of the physical
$B^0_d$--$\bar B^0_d$--mixing eigenstates. Note that eq.~(\ref{e9}) is
only valid in the case of $B_d$-decays into final CP-eigenstates $|f\rangle$
satisfying $({\cal CP})|f\rangle=\pm|f\rangle$. This requirement is
fulfilled by the $B_d$-modes considered in this letter. Let us note that
we would have $\psi=0$ if we neglected the QCD-penguins with internal
up- and charm-quark exchanges in the mode $\bkk$, and $\phi=2\alpha$ if we
omitted the penguin contributions to the decay $\bpipi$.

The knowledge of $\psi$ and $\phi$ together with the branching ratios
for the decays $\bdkk$ and $\bdpipi$ and their CP-conjugates,
respectively, specifies the dashed and solid triangles shown in Fig.~1.
If we knew in addition the angle $\sigma$ fixing the relative orientation
of these two triangles, the angle $\alpha'$ in Fig.~1 could be determined.
It is related to the CKM-angle $\alpha$ through
\beq\lab{e10a}
\alpha=\alpha'+\delta\alpha,
\eeq
where
\beq\lab{e10b}
\delta\alpha=\order\left(\frac{1}{2}\frac{|P_3|+|\bar P_3|}{|T+E|}\right)=
\order(\bar\lambda^2).
\eeq
Note that $\delta\alpha$ is of the same order in $\bar\lambda$ as the
uncertainty affecting the determination of $\alpha$ by using the
$B\to\pi\pi$ approach proposed by Gronau and London \cite{gl,ghlrewp}.
In contrast to eq.~(\ref{e10b}), the latter uncertainty is not related to
$SU(3)$-breaking effects but originates from electroweak penguin
operators.

While the angles $\psi$ and $\phi$ are measured by the CP-violating
asymmetries (\ref{e7}) and (\ref{e8}), respectively, the angle $\sigma$
can only be determined in an {\it indirect} way. To this end, let us
neglect the $SU(3)$-breaking effects described by the amplitudes $P_3$
and $\bar P_3$ which are both $\order(\bar\lambda^2)$.
If we define the quantities
\beq\lab{e11}
\begin{array}{rclrcl}
A&\equiv&|A(\bdpipi)|,&\bar A&\equiv&|A(\bdbpipi)|,\\
B&\equiv&|A(\bdkk)|,&\bar B&\equiv&|A(\bdbkk)|,
\end{array}
\eeq
we obtain in this strict $SU(3)$-symmetric case the following
equations from Fig.~1:
\bea
A^2+B^2-2AB\cos(\psi+\sigma)&=&|T+E|^2\lab{e12}\\
\bar A^2+\bar B^2-2\bar A\bar B\cos(\sigma+\phi)&=&|T+E|^2.\lab{e13}
\eea
Combining (\ref{e12}) and (\ref{e13}) yields the equation
\beq\lab{e14}
a\cos\sigma-b\sin\sigma=c,
\eeq
where we have introduced the quantities $a$, $b$ and $c$ through
\beq\lab{e15a}
a\equiv\bar A\bar B\cos\phi-AB\cos\psi
\eeq
\beq\lab{e15b}
b\equiv\bar A\bar B\sin\phi-AB\sin\psi
\eeq
\beq\lab{e15c}
c\equiv\frac{1}{2}\left(\bar A^2+\bar B^2-A^2-B^2\right).
\eeq
Its solution can be written in the form
\beq\lab{e14a}
\tan\sigma=\frac{-bc\pm a\sqrt{a^2+b^2-c^2}}{ac\pm b\sqrt{a^2+b^2-c^2}}
\eeq
and fixes $\tan\sigma$ up to a two-fold ambiguity corresponding to ``$+$''
and ``$-$'', respectively. Consequently, $\sigma$ can be determined up to
a four-fold ambiguity. Note that there would be no ambiguity in
$\tan\sigma$ in the special cases $c=0$, which corresponds to the limit
of no direct CP violation in the decays $\bpipi$ and $\bkk$, and
$a^2+b^2-c^2=0$. The angles $\psi$ and $\phi$ determined by using
(\ref{e7}) and (\ref{e8}), respectively, suffer also from two-fold
ambiguities which are a characteristic feature
of the determination of angles by using CP-violating asymmetries
or amplitude relations. Taking into account additional information
from other processes, it should be possible to exclude certain solutions
and to resolve these ambiguities. In particular the future knowledge of
the shape of the unitarity triangle obtained from loop induced
transitions (see e.g.~\cite{blo}) should be useful in this respect.

Using $\sigma$ determined by means of eq.~(\ref{e14a}), both the
angle $\alpha$ and the quantity $|T+E|$ can be extracted in the limit of
vanishing $SU(3)$-breaking, i.e.\ $P_3=\bar P_3=0$, as can be seen from
Fig.~1. One could easily generalize the equations above by including the
effect of $P_3$ and $\bar P_3$. This would modify $\sigma$ and
consequently $\alpha$ by corrections of $\order(\bar\lambda^2)$. Due to
the lack of knowledge of the exact values of $P_3$ and $\bar P_3$ this
generalization would not improve the accuracy of our method at present.

Consequently, combining all these considerations
(see also eq.~(\ref{e10b})), we expect the
uncertainty in the determination of $\alpha$  in our approach
to be of $\order(\bar\lambda^2)$. It should be
stressed -- as has already been done in refs.~\cite{ghlrewp,ghlrsu3} --
that this estimate should not be taken too literally since $\bar\lambda
=\order(0.2)$ is not a small number. Therefore, in practice the accuracy
of our approach may well be of $\order(\bar\lambda^{2\pm1})$. In order to
control it in a quantitative way, we have to deal with the $SU(3)$-breaking
contributions $P_3$ and $\bar P_3$ which is beyond the scope of this
letter. In this respect the $\order(\bar\lambda^2)$
electroweak penguin uncertainty affecting the determination of
$\alpha$ in the Gronau-London-method~\cite{gl}
is in better shape as we have shown in ref.~\cite{bfewp}.
Performing measurements of the branching ratios of certain $B\to\pi K$
channels, which are expected to be
of $\order(10^{-5})$, these electroweak penguin effects can be determined
in principle.

In summary we have presented a determination of the CKM-angle $\alpha$
by using mixing-induced CP violation in the decays $\bpipi$ and
$\bkk$. Interestingly enough, the accuracy of our method, which is
limited by $SU(3)$-breaking effects related to the creation of $s\bar s$
pairs in $\bar b\to\bar d$ QCD-penguin processes, is expected to be of
the same order in $\bar\lambda$, i.e.\ $\order(\bar\lambda^2)$, as the
one arising from electroweak penguins in the original $B\to\pi\pi$
approach of Gronau and London. As we stated above, the electroweak
penguin uncertainties in the latter method can be brought under control as
demonstrated in ref.~\cite{bfewp}, whereas this is not the case of the
$\order(\bar\lambda^2)$ $SU(3)$-breaking effects present in the method
described here. Despite of this our method may be an interesting
alternative to determine the CKM-angle $\alpha$ in a rather clean way.
An advantage of our approach is the fact that it does not involve
a measurement of the decay $B_d\to\pi^0\pi^0$ which is considered to be
difficult. However, we need instead a time-dependent analysis
of the pure penguin-induced mode $B_d\to K^0\bar K^0$. Experimentalists
will find out which method can be performed easier in practice. It is
needless to say that a comparison of $\alpha$ determinations by means
of these two methods would give another test of the CKM picture
of CP violation.

\vspace{0.5cm}

A.J.B. would like to thank Iris Abt for illuminating discussions.

\newpage

\section*{Figure Caption}

\vspace*{1cm}

\begin{table}[h]
\begin{tabular}{ll}
Fig.\ 1:&A different strategy for determining the CKM-angle $\alpha$.\\
&\\
\end{tabular}
\end{table}

\newpage

\begin{figure}[p]
\vspace{15cm}
\caption{}\lab{f1}
\end{figure}

\end{document}